# Enhancing Passive Non-Line-of-Sight Imaging Using Polarization Cues

Kenichiro Tanaka*†    Mukaigawa Yasuhiro*    Achuta Kadambi†
*NAIST, Japan    †UCLA, USA
ktanaka@is.naist.jp    mukaigawa@is.naist.jp    achuta@ee.ucla.edu

## Abstract

*This paper presents a method of passive non-line-of-sight (NLOS) imaging using polarization cues. A key observation is that the oblique light has a different polarimetric signal. It turns out this effect is due to the polarization axis rotation, a phenomena which can be used to better condition the light transport matrix for non-line-of-sight imaging. Our analysis and results show that the use of a polarizer in front of the camera is not only a separate technique, but it can be seen as an enhancement technique for more advanced forms of passive NLOS imaging. For example, this paper shows that polarization can enhance passive NLOS imaging both with and without occluders. In all tested cases, despite the light attenuation from polarization optics, recovery of the occluded images is improved.*

## 1. Introduction

Non-Line-of-Sight (NLOS) imaging is a very active research topic in the field of computational imaging. The goal is to visualize a scene that is hidden from the camera's line of sight, *e.g.*, "looking around the corners". Several prior works have tackled this problem, using methods that range from (a) time of flight imaging [40, 18, 26, 30, 25, 28, 37, 6, 14, 10, 41, 38, 29, 2]; (b) wave optics [21, 11, 10, 41]; (c) shadows [5, 36, 42, 3, 35]; and even (d) machine learning [34, 9, 8]. This paper takes a different tack, proposing the use of polarization cues to re-examine the NLOS problem.

Of particular interest is passive NLOS imaging where one is unable to control the probing illumination. Such limited programmable control makes scene reconstruction very challenging—existing passive NLOS methods [32] offer blurry reconstructions, as compared to active NLOS. This "blur" in existing passive NLOS methods can be mathematically linked to the scene's light transport matrix. To obtain better recovery, previous work aims to reduce the condition number of the light transport matrix. This has been done, for example, by placing a partial occluder in the scene to create high-frequency shadows [32].

Our method is analogous to prior approaches in passive NLOS, but we make a first attempt to use (linear) polarization cues to improve the conditioning of the light transport matrix. Our method creates high-frequency variation in the light transport matrix by, ideally placing the camera at the Brewster angle of polarization. In this oblique angle, the polarizer's axis varies and changes the intensity depending on the oblique angle. We refer this effect as *effective angle* of polarizer. We show in the paper that this oblique observation provably changes the conditioning of the light transport matrix. Further, these benefits of polarization can apply to multiple configurations for passive NLOS. For example, polarization can be used to enhance occluder-based passive NLOS or direct passive NLOS imaging.

In summary, we make the following contributions:

- We bring the polarizer's effective angle theory to the computer vision field. The polarizer's axis depends on both zenith and azimuth angle of the light ray, which conveys rich angular information;

- We demonstrate that the polarization cues are able to improve the conditioning of the light transport for passive NLOS imaging without scene modifications; and

- We demonstrate that the same polarization cues also improve other passive NLOS approaches, including those that use partial occluders.

**Scope:** While polarization is a fresh signal for use in NLOS imaging, the quality of passive NLOS (after polarization enhancement) does not approach that of active methods, which have been shown to obtain extremely high-fidelity reconstructions. However, our polarization enhancement is fundamentally more general than the results we present here. A future extension details how the proposed technique could apply to active NLOS imaging, covered at the end of this paper as the appendix.

## 2. Related Work

In this section, we briefly review the related work regarding NLOS imaging. For a more comprehensive review of NLOS imaging, the readers are directed to [22].

**Active NLOS imaging.** NLOS imaging was first proposed in the context of active, time-resolved imaging by Raskar and Davis [31]. Later work experimentally demonstrated and theoretically evolved these ideas through the use of time of flight imagers, in particular streak cameras [40, 39, 18, 26], amplitude-modulated continuous-wave cameras [13, 16, 15], and single photon avalanche diodes (SPAD) and SPAD cameras [30, 25, 28, 37, 6, 14, 10, 41, 38, 29, 2]. There are other methods of performing active NLOS imaging that do not require time resolved information. For example, a coherent light source reveals occluded cues [4]. An object movement can also be tracked by speckle [33] or synthesis-based approaches [19]. Recent work has used a standard RGB camera and laser source to realize active NLOS [9]. While active illumination increases the scene information, we choose to focus on enhancing the lower-performing, but more flexible configuration of passive NLOS imaging.

**Passive NLOS imaging** There are fewer works that study the hard problem of passive NLOS. One promising approach is to use shadows and corners. Bouman *et al.* [5] use the high-frequency detail of a corner to track occluded scenes (this work is partially inspired by accidental pinhole cameras [36]). An extension of this is proposed in [32, 42, 12], where a partial obstacle is placed between the wall and NLOS scene. An orthogonal approach is to image thermal scenes around the corner. Here, heat is passively emitted by the human body, which simplifies the NLOS problem to a 1-bounce reflection, enabling high-quality video of a human figure, in real-time [23, 17]. Our technique is complementary, as polarization can enhance the quality of most methods referenced above.

**Analysis of NLOS imaging** The recoverability of NLOS imaging depends on many factors. Kadambi *et al.* [16] propose the first bound on the spatial resolution of NLOS imaging, and in particular active NLOS. Liu *et al.* [20] analyze the feature visibility of SPAD-based NLOS imaging. Saunders *et al.* [32] analyze the aperture of NLOS imaging using the size of the LOS wall and the obstacles. Pediredla *et al.* [29] propose a temporal focusing using ellipsoidal projection. In this paper, we follow the structure of previous techniques in analyzing how the condition number of the light transport matrix is favorably modified through the use of polarization cues.

**Effective angle of polarizer** In the LCD development field, the light leakage of polarizer from oblique view is a major problem [43, 27]. While they aim to cancel this effect, we bring this effect to improve the NLOS imaging problem.

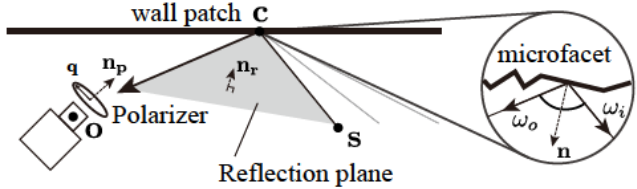

Figure 1: **Diagram of the geometry of the problem.** The camera is looking at the wall and the scene point is out of sight of the camera. On a microfacet model, each reflection path can be considered as a sum of mirror reflections, hence the polarization is preserved.

## 3. Light Transport: Passive NLOS Imaging

Suppose a camera is pointed toward the line-of-sight (LOS) wall and the scene is non-line-of-sight (NLOS) to the camera as shown in Fig. 1. Without loss of generality, we consider that the NLOS scene is a set of point light sources. Then, the intensity $I(\mathbf{c})$ at the wall patch c is given by

$$I(\mathbf{c}) = \iint_{\mathbf{s} \in \mathcal{S}} T(\mathbf{s}, \mathbf{c}) l(\mathbf{s}) d\mathbf{s}, \quad (1)$$

where s is a point of in the scene $\mathcal{S}$, $l(\mathbf{s})$ is the intensity of the scene point s, and $T(\mathbf{s}, \mathbf{c})$ is the light transport from the scene point s to the camera via the wall patch c. It is able to discretize and superpose the observation, such that

$$\mathbf{i} = \mathbf{T}\mathbf{l}, \quad (2)$$

where i is the vectorized observations, **T** is the light transport matrix, and l is the vectorized scene intensities. If the light transport matrix is known or generatable, the NLOS intensities can be estimated by least squares sense as

$$\hat{\mathbf{l}} = \mathbf{T}^+ \mathbf{i}, \quad (3)$$

where $\mathbf{T}^+$ is the pseudo-inverse matrix of **T**. The stability of solving this linear system depends on how small the condition number of the matrix is. A key goal of previous methods has been to improve the conditioning of Eq. (3).

**Previous methods of conditioning Eq. (3):** Perhaps the simplest way to decrease the condition number is to use a favorable bidirectional reflectance distribution function (BRDF). A trivial case is the mirror, which makes the lowest condition number because the light transport matrix becomes identity. An opposite example is the diffuse wall, which makes a very large condition number because the single light source contributes to the all camera pixels. The conditioning of using other materials that have specularity are between the mirror and the diffuse wall because they somehow preserve high frequency component. Extended discussion on this topic can be found in [16]. Another way to improve the conditioning is to place obstacles in the scene, such as putting a partial obstacle between the wall and the scene. Saunders et al. [32] place an arbitrary obstacle between the camera and the wall to block the light rays. This makes a shadow on the wall, which contains high-frequency information. For more detail, the readers are referred to papers that use obstacles [32, 42, 12]. Both of these approaches modify the scene.

## 4. Light Transport: Polarized NLOS Imaging

We aim to minimize modifications to the scene for conditioning Equation 3 using polarization. Hence, we only use a polarizer at the camera side. This approach is also able to conditioning the existing method using partial occluders. If putting a partial occluder to the scene is tolerable, combining the existing method and the proposed method improves the conditioning further.

By putting a polarizer in front of the camera, a small angular difference of light paths makes a big intensity variance, thus the conditioning is improved compared to a normal observation without polarizer. A key observation of this paper is that the polarizer's effective axis is slightly rotated if the light ray is oblique to the polarizer. In other words, the perpendicular light rays are blocked while oblique light rays pass through the polarizer. In the following section, we reveal how the polarization light transport is modeled in passive NLOS imaging.

**Polarized NLOS scenes** If the NLOS scene itself is polarized, then we can exploit cross-polarization effects to improve the NLOS problem even more. Further detail is deferred to the supplementary material.

### 4.1. Effective angle of polarizer

We introduce the effective angle of a polarizer, which is well studied in the area of LCD development [43, 27]. When the light ray is oblique to the polarizer, the light is 'leaked' even if two linear polarizers are put crossed. This is because the effective polarization axis of the polarizer depends on the azimuth and zenith angle of the incident light ray. Figure 2(a, b) shows pictures of the same scene from top and oblique views. While the light from the LCD is blocked on the top view, the content of the LCD is slightly visible from oblique view even though the polarizer is put crossed. This effect is angle dependent, therefore it can be used for analyzing NLOS observations.

Light leakage occurs because the effective angle of the polarizer changes due to the light ray's azimuth and zenith angles. From a geometrical calculation[1], the effective angle $\theta'$ from the incident light viewpoint is represented as

$$\tan \theta' = -\frac{\cos(z)}{\tan(\theta - a)}, \quad (4)$$

where $\theta$ is the original polarizer's axis, i.e., the polarizer's axis from top view, $a$ and $z$ are azimuth and zenith angles of the incident ray, respectively. Figure 2(c - e) shows the original and effective angles of polarizers and the light leakage pattern of crossed polarizers. This pattern can be utilized to improve the NLOS imaging.

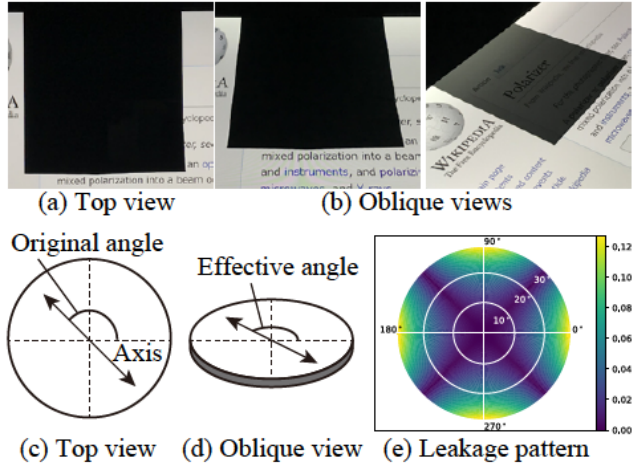

(a) Top view  (b) Oblique views

(c) Top view  (d) Oblique view  (e) Leakage pattern

Figure 2: **Polarizer from oblique view.** While the LCD monitor is invisible from top view (a), it is slightly visible from certain oblique views depending on the zenith and azimuth angles (b). (c) The original angle of polarizer from top view. (d) The effective angle from oblique view. The polarizer axis is slightly declined. (e) Light leakage pattern of crossed polarizers from oblique view. These polarizers' original angles are $45°$ and $-45°$, respectively.

### 4.2. Polarization light transport on rough surface

Now, we consider the polarization light transport model on a rough surface. In this paper, we employ a microfacet model for the rough surface, shown in the inset of Fig. 1. The surface normal of each facet that reflects the light source to the camera is identical to the half vector of the viewing and lighting vectors. Therefore, the light transport $T(\mathbf{s}, \mathbf{c})$ from the scene point $\mathbf{s}$ to the camera via the wall patch $\mathbf{c}$ is represented as

$$T(\mathbf{s}, \mathbf{c}) = \Omega(\omega_\mathbf{i}, \omega_\mathbf{o}) \lambda(\omega_\mathbf{i}, \omega_\mathbf{o}, \mathbf{q}), \quad (5)$$

$$\begin{cases} \omega_\mathbf{i} = \frac{\mathbf{s}-\mathbf{c}}{\|\mathbf{s}-\mathbf{c}\|_2}, \\ \omega_\mathbf{o} = \frac{\mathbf{o}-\mathbf{c}}{\|\mathbf{o}-\mathbf{c}\|_2}, \end{cases}$$

---
[1]Refer supplementary for the detail.

where $\Omega$ is the BRDF of the rough surface, $\omega_o$ is the viewing vector, $\omega_i$ is the incident vector, o is the camera position, $\lambda$ is the light leaking/blocking effect due to polarizer, and q is the polarizer's axis. The polarizer modulates light transport by introducing the leakage term $\lambda$, which makes the improvement to the light transport matrix.

As we assume the microfacet model, the reflection on each facet can be modelled as a Fresnel reflection. The Fresnel reflection is known to be partially polarized and reflectances of s and p polarization components $R_s$ and $R_p$ can be represented as

$$R_p(\phi) = \frac{\tan^2(\phi - \phi')}{\tan^2(\phi + \phi')}, \quad (6)$$

$$R_s(\phi) = \frac{\sin^2(\phi - \phi')}{\sin^2(\phi + \phi')}, \quad (7)$$

$$\phi' = \sin^{-1}\frac{\sin(\phi)}{\eta}, \quad (8)$$

where $\phi$ is the incident angle, $\phi'$ is the refractive angle, and $\eta$ is the refractive index of the wall.

When the incident and reflection angle is at Brewster angle, the reflected light is completely linearly polarized. This is because the reflectance of p polarization becomes zero. By putting the camera at near the Brewster angle position as shown in Fig. 3, highly polarized observations can be obtained, thus the observation can be analyzed with a polarizer in front of the camera.

Placing a polarizer in front of the camera such that the polarized reflection is blocked at a specific light path, leads to the observation of a light leakage pattern. Because the other light paths from neighboring wall points are oblique to the polarizer's blocking axis, the leakage pattern can be observed as shown in the right of Fig. 3.

The leakage pattern $\lambda$ can be modeled as

$$\lambda(\omega_i, \omega_o, q) = R_p(\theta_h)\cos(\theta') + R_s(\theta_h)\sin(\theta'), \quad (9)$$

$$\begin{cases} \theta_h &= \frac{1}{2}\cos^{-1}(\omega_o \cdot \omega_i) \\ \theta' &= \tan^{-1}\left(-\frac{\cos(z)}{\tan(\theta-a)}\right) \\ z &= \cos^{-1}(-\omega_o n_p) \\ \theta - a &= \cos^{-1}(-wq) \\ w &= \frac{\omega_o + \cos(z)n_p}{\|\omega_o + \cos(z)n_p\|_2} \end{cases}$$

where q is the polarizer's axis, $\theta_h$ is the half angle of the reflection path, $\theta'$ is the effective angle of the polarizer, and $n_p$ is the normal of the polarizer.

**Combination with the existing method** Polarization light transport can be combined with existing method such as putting partial occluder in the scene [32]. The light transport matrix in this case becomes

$$T(c, s) = T'(c, s)\lambda(\omega_i, \omega_o, q), \quad (10)$$

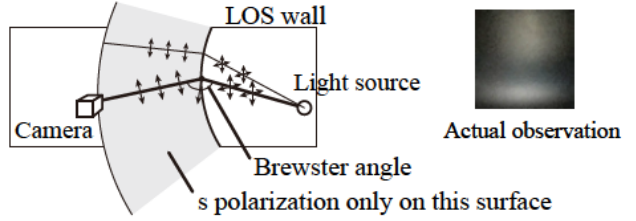

Figure 3: **The Brewster angle geometry.** Putting the camera so that light path is at Brewster angle, only one directional polarization (s polarization component) is reflected to the camera. Because there is no effect from p polarization, it is possible to observe a light leakage pattern only using a polarizer in front of the camera.

where $T'$ is the light transport matrix of the existing method. Again, the difference is the existence of $\lambda$ and this improves the condition number of the light transport matrix.

### 4.3. Other factors

**Polarized scene** Although most NLOS scenes are unpolarized, in the rare cases where the NLOS scene is polarized, our method is extremely advantageous. Consider that if the scene is polarized such as an LCD monitor, Eq. (9) can be rewritten as

$$\lambda(\omega_i, \omega_o, q) = I_p R_p(\theta_h)\cos(\theta') + I_s R_s(\theta_h)\sin(\theta'), \quad (11)$$

where $I_p$ and $I_s$ is the intensity of p and s polarization components of the scene. Here, it is possible to place the polarizer at an angle where one of the polarization components becomes zero. Therefore, there is no restriction of having the capture setup oriented to the Brewster angle, as in the general case we have described above. We expand on this discussion in the supplement, and show results.

**Why not rotate the polarizer?** This paper relies on taking 1 image from a polarization filter at 90 degrees (parallel to the reflection plane). A natural question is whether the filter can be rotated to take multiple pictures. Since we rely on a leakage pattern, there is little benefit to capturing multiple images, as the variation in polarization images at angles other than 90 degrees is subtle, while the capture effort increases linearly. Very specifically, the key improvement from using polarizers comes from the 'dark band' in the image as shown in Fig. 4. Figure 4 shows the observation image while rotating the polarizer and without the polarizer. The dark band, where the light is blocked, appears only around 90 degrees, while other angles look similar to the image without the polarizer.

**Relationship to AoLP/DoLP** The normalized 90 degree image in Fig. 4 is compatible to the angle of linear polar-ization (AoLP) and degree of linear polarization (DoLP) if the scene is only a single point light source. The benefit of using an intensity image is that it is linearly addable and requires only a single image at least. On the other hand, AoLP and DoLP requires a tricky trigonometric calculation and at least 4 polarization rotations.

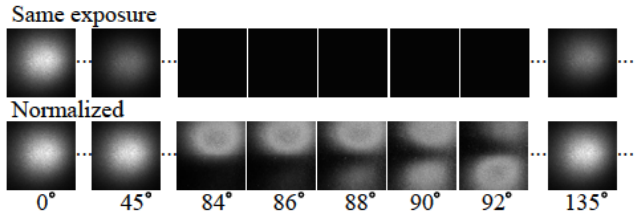

Figure 4: **Comparison to rotating the polarizer.** The upper row is the captured images of the same exposure time, and the lower images are normalized at each maximum value. The dark band, where the light is completely blocked by polarizer, is a key observation and only appears around 90 degrees.

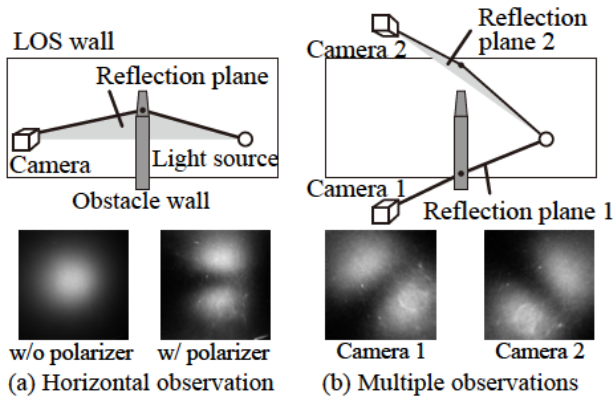

Figure 5: **Camera position and reflection plane.** (a) Using a single camera position, only one direction is encoded in the image. The intensity variance of horizontal direction is not so improved. (b) Using multiple camera positions, the scene can be encoded by multiple directions.

**Multiple camera positions** The leakage pattern appears parallel to the reflection plane. Figure 5(a) shows the actual measurement of the wall, where the scene is a point light source. There is large intensity variance in the vertical direction, and therefore the vertical information is better preserved. On the other hand, the horizontal intensity variance is lower and likely to the unpolarized observations. To overcome the slim variance in the horizontal direction, it is possible to capture the scene from multiple camera positions, enabling the capture of perpendicular reflection planes as shown in Fig. 5(b).

|  | **Standalone** | | | |
|---|---|---|---|---|
|  | w/o | rotating | single | **multiple** |
| Cond. num. | 686.8 | 486.3 | 357.0 | **327.9** |
| Percentage | – | 70.8% | 52.0% | **47.7%** |
|  | **With partial occluder** | | | |
|  | w/o | rotating | single | **multiple** |
| Cond. num. | 172.3 | 170.6 | 146.1 | **113.9** |
| Percentage | – | 99.0% | 84.8% | **66.1%** |

Table 1: **Condition number comparison.** Methods without polarizer, rotating polarizer, crossed polarizer from a single camera position, and crossed polarizer from multiple camera positions are compared. Our method has the lowest condition number.

## 5. Simulation

Our simulations verify that polarization leakage can be used to improve the condition number of the light transport matrix.

**Condition number** The effectiveness of the method is confirmed by examining the condition number of light transport matrices. The lower condition number gives better recovery quality. The camera and the light source are placed 10 cm from the wall and the scene is assumed to be a set of $3 \times 3$ point light sources. For a single image observation, the camera position A is selected, and for multiple camera settings, positions B and C are used. The condition number for each setting is summarized in Table 1. Our method has the smallest condition number, indicating the potential for about two times better recoverability. Likewise, we also show the case when there is a partial occluder in the same table. We observe through simulation that the condition number is improved using a polarizer as well.

**Improvement w.r.t. wall roughness** The effectiveness of this method depends on the type of the wall. To confirm this, the improvement in condition number is evaluated while changing the roughness of the wall. The roughness is changed from 0 (mirror-like) to 1 (completely diffuse). The plot of the condition number and improvement ratio is shown in Fig. 6. For all roughness parameters, it is observed that the polarization cues improve the condition number. Although the plots of single and multiple cameras look similar, we observe that the multiple camera setting is always slightly better than single camera setting. When the roughness decreases, the wall becomes mirror-like and there is no improvement using a polarizer because the NLOS scene is originally visible. The best performance is observed at the middle of specular and diffuse reflections, where a lot of materials have such specularities [24].

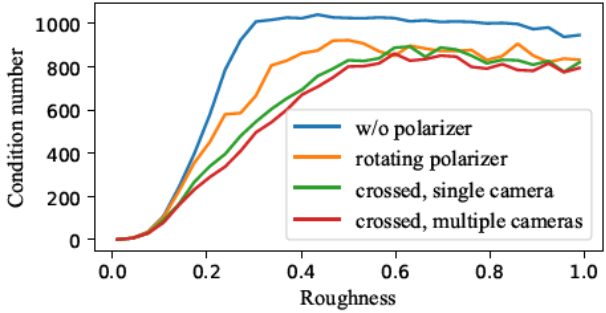

Figure 6: **Condition number with respect to the roughness of the wall.** Lower the better. The roughness ranges 0 (mirror-like) to 1 (diffuse-like). Our method has the lowest condition number for all roughness parameter.

## 6. Experiment

Real experiments are consistent with simulations: polarization improves NLOS image reconstruction. For all following experiments, we use ADMM to solve Eq. (2), consisting of a 2D total variation regularizer with a box constraint. For clarity, we estimate

$$\hat{\mathbf{l}} = \underset{\mathbf{l}}{\arg\min} \|\mathbf{i} - \mathbf{T}\mathbf{l}\|_2^2 + \lambda TV_{2D}(\mathbf{l}) \quad (12)$$
$$\text{s. t.} \quad 0 \preceq \mathbf{l} \preceq 1.$$

Because this is a convex optimization problem, it can be solved in a polynomial time. For all cases, the BRDF of the wall is measured beforehand.

**Polarized NLOS** Firstly, we evaluate the polarized NLOS without partial occluder. For numerical evaluation of non-polarized scene, a projector is used to project the scene image. Figure 7 shows the setup and the result. Two images are compared with and without the polarizer in front of the camera. While it is difficult to see the projected scenes if the polarizer is not used, the scene is visible using the polarizer. We also projected more images, which can be found in the supplementary material. The table shows the numerical evaluation of the results. Peak signal to noise ration (PSNR), zero-mean normalized cross correlation (ZNCC), and structural similarity (SSIM) are used. It is confirmed that the condition number is decreased and the recovered image is improved for every image metric if the polarization is used.

**Polarized NLOS w/ partial occluder for reflective objects** Here, we show that Polarized NLOS can also enhance existing techniques. As shown in Fig. 8 *top-left*, we reproduce the partial occluder method from Saunders *et*

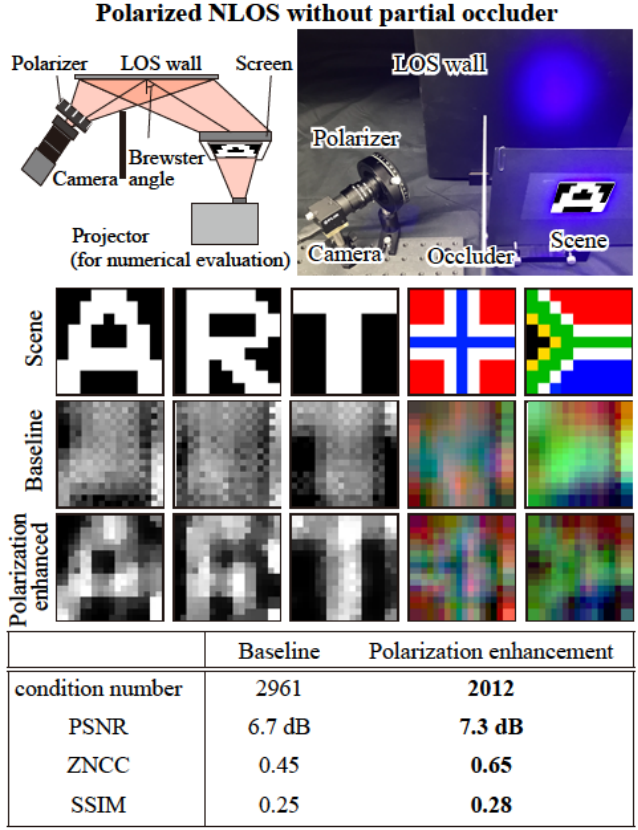

Figure 7: **Polarized NLOS results without occluder.** A projector is used for numerical evaluation and to make the scene unpolarized. The scene is recovered with and without the polarizer in front of the camera. Using polarization, the recovered images are improved. Improvement is confirmed by comparing condition number and three image measures.

*al.* [32]. Reflective object scenes are recovered in this experiment and enhanced by polarization. Figure 8 shows the setup, the target object, the recovered result by the existing method, and the enhanced result by polarization. The target object is lit by an uncontrolled light source. In the results of the baseline method, it is difficult to see the resolution chart and the content of the book. On the other hand, our technique recovers images with higher contrast. The clear texture of resolution chart and printed materials are visualized in detail. Our method quantitatively and qualitatively improves the reconstruction. We provide more diverse scenes in the supplement.

**Comparing our enhancement to image processing** An interesting question that is raised is whether the performance improvements we obtain could be achieved by applying image post-processing algorithms to conventional NLOS (without polarization). Figure 9 shows the result of projected images and also includes the result of applying image post-processing. The baseline method in this case is [32]. Here, a total variation (TV) denoising algorithm [7] and a deep learning image processor (neural enhance) [1] are used. While the image is improved by post processing, it is impossible to recover the higher frequency component that is lost on the wall reflection, because recovering lost information is mathematically impossible. On the other hand, our method recovers higher frequency detail, which is preserved by polarization light transport. PSNR, ZNCC, and SSIM values show significant improvement.

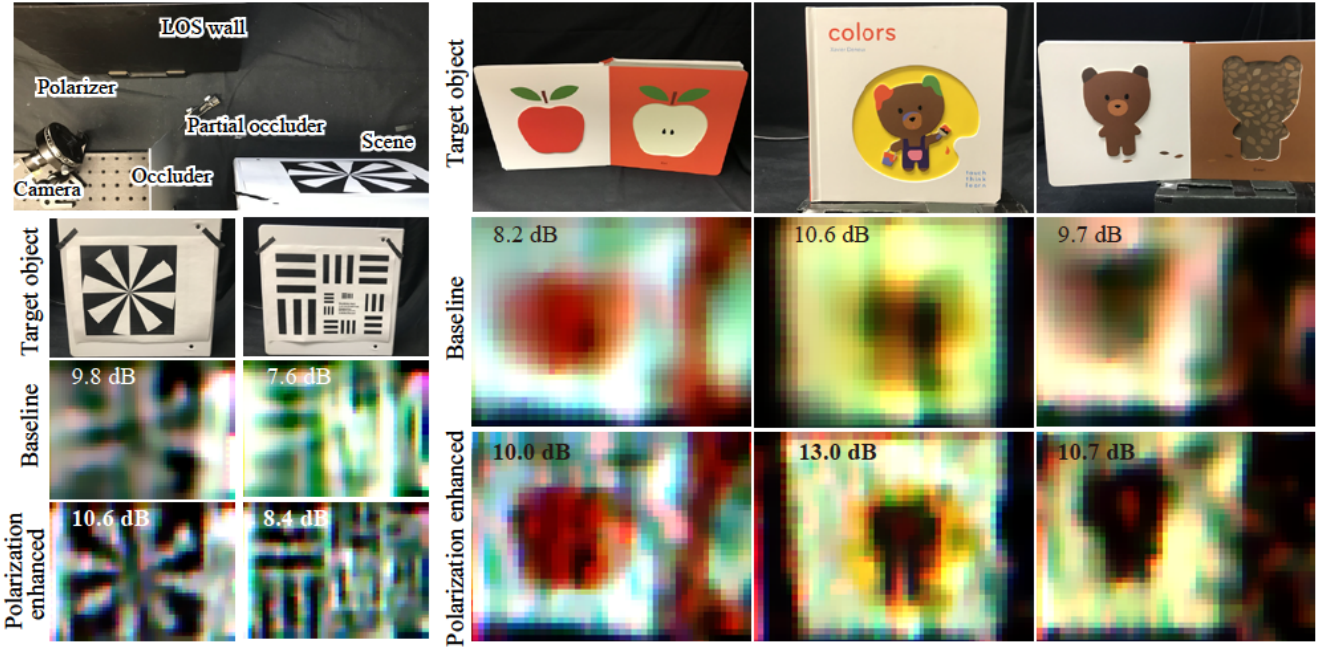

Figure 8: **Results for reflective objects.** *Top-left:* the setting of the experiment. The scene is a reflective object (not self-luminous). *1st row:* The photograph of target objects. *2nd row:* The recovered images by the baseline method [32]. *Bottom row:* The recovered images by our method. High frequency details are recovered. Clear detail of resolution charts, sharp edge of apple, and the detailed shape of bears are clearly visualized. PSNR values are calculated with homography-transformed photograph for reference.

## 7. Discussion

In this paper, we propose to use polarizer for conditioning the passive NLOS imaging. Throughout our experiment, we confirm that the our method improves the conditioning for general passive NLOS imaging scenarios. As this approach optically modifies the light transport, it is mathematically different from any other post image processing algorithm based on lower frequency observation. The results of our method are better than TV denoising and neural image enhancement.

It is not very sensitive to the Brewster angle position. This is because the polarization component varies due to Eq. (9) while the most effective alternation in intensity occurs at the Brewster angle. Indeed, our experiments are performed by putting the camera roughly but not exactly at the Brewster angle. It is expected that a part of the light path eventually meets the Brewster angle geometry.

As we assume the polarizers axis is around the blocking angle, the image is dark and requires a longer exposure. Because the polarization information theoretically exists at any polarization angle, it is possible to use another polarization angle. However, although a shorter exposure can be used, the ratio of the polarization signal to the mean brightness gets worse so it is not realistic. This is the tradeoff of the sensor's dynamic range and the optical signal-to-bias ratio.

Our method also assumes that the wall preserves the polarization property. This holds for many rough surfaces, but it is generally a subtle signal. The wall that has dominant subsurface scattering, such as plastic and plaster, is very challenging because it loses the polarization property and the ratio of polarization due to Fresnel reflection becomes low. Possibility of capturing such weak signal depends on the dynamic range of the sensor. Alternatively, a combination of rotating a polarizer and an event camera that can capture relative intensity changes can be utilized, which is an interesting future direction of capturing weak polarization signals.

Our results support the idea of using polarization cues for NLOS light transport analysis. Although we only apply the idea to the passive NLOS experiments in this paper, the proposed method can be extended to active NLOS imaging as shown in the appendix. In conclusion, we hope this paper spurs interest in using polarization for problems in NLOS imaging and multipath light transport.

## Appendix: On Active NLOS Imaging

Followed by the previous work [2, 13, 28, 18, 40], an active NLOS imaging using time-of-flight measurement can be modeled as

$$i(t; \mathbf{p}, \mathbf{c}) = \iiint_{\mathbf{s}\in\mathcal{S}} \rho(\mathbf{s}) \frac{\delta(\|\mathbf{s}-\mathbf{p}\| + \|\mathbf{s}-\mathbf{c}\| - ct)}{\|\mathbf{s}-\mathbf{p}\|^2 \|\mathbf{s}-\mathbf{c}\|^2} \times$$
$$\Omega(\omega_{il}, \omega_{ol})\Omega(\omega_{ic}, \omega_{oc}) d\mathbf{s}, \quad (13)$$

$$\begin{cases} \omega_{il} &= \frac{\mathbf{s}-\mathbf{p}}{\|\mathbf{s}-\mathbf{p}\|_2}, \\ \omega_{ol} &= \frac{\mathbf{o}-\mathbf{p}}{\|\mathbf{o}-\mathbf{p}\|_2}, \\ \omega_{ic} &= \frac{\mathbf{s}-\mathbf{c}}{\|\mathbf{s}-\mathbf{c}\|_2}, \\ \omega_{oc} &= \frac{\mathbf{o}-\mathbf{c}}{\|\mathbf{o}-\mathbf{c}\|_2}, \end{cases}$$

where $i(t; \mathbf{p}, \mathbf{c})$ is the temporal transient observation at the wall patch $\mathbf{c}$ by illuminating the wall patch $\mathbf{p}$ by a pulsed light, $\rho$ is the 3D NLOS albedo, $\delta$ is the Dirac delta function, $c$ is the speed of light, and $\Omega$ is the BRDF of the LOS wall. Discretizing, we obtain

$$\mathbf{i} = \mathbf{T}\rho, \quad (14)$$

where $\mathbf{T}$ is the light transport matrix of active NLOS model and $\rho$ is the vectorized NLOS albedo.

Analogous to the passive case, when we put a polarizer in front of the camera, Eq. (13) is altered as

$$i'(t; \mathbf{p}, \mathbf{c}) = \iiint_{\mathbf{s}\in\mathcal{S}} \rho(\mathbf{s}) \frac{\delta(\|\mathbf{s}-\mathbf{p}\| + \|\mathbf{s}-\mathbf{c}\| - ct)}{\|\mathbf{s}-\mathbf{p}\|^2 \|\mathbf{s}-\mathbf{c}\|^2} \times$$
$$\Omega(\omega_{il}, \omega_{ol})\Omega(\omega_{ic}, \omega_{oc})\lambda(\omega_{ic}, \omega_{oc}, \mathbf{q}) d\mathbf{s}, \quad (15)$$

where the light transport is modulated by the leakage term $\lambda$. Note that the polarization state of active light source is ignored because the light is assumed to be depolarized at the NLOS scene reflection.

**Simulation** We confirm the effectiveness of polarized NLOS for active NLOS imaging through a simulation. Figure 10 shows the condition number of the light transport matrix changing with scene configurations including wall's roughness parameter and the spatial resolution of NLOS scene. For all configurations, polarization cues improve the conditioning of the active NLOS imaging.

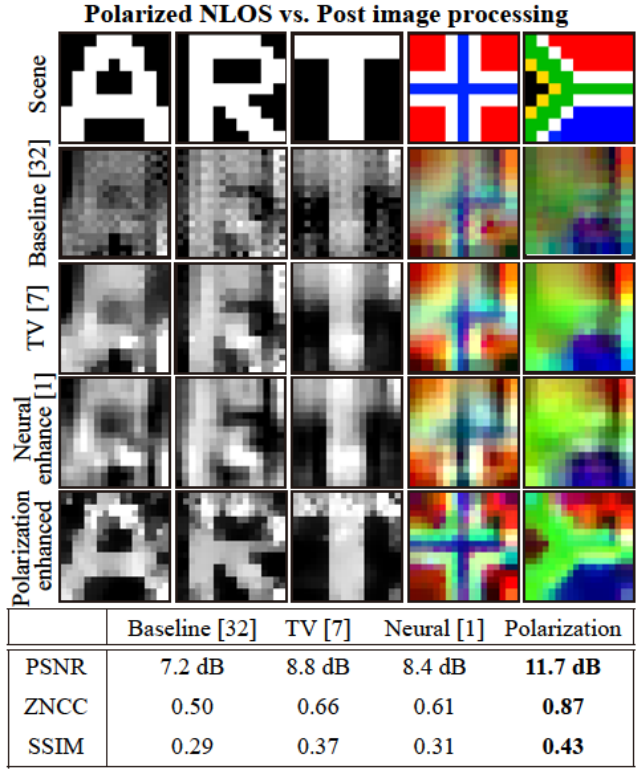

|  | Baseline [32] | TV [7] | Neural [1] | Polarization |
|---|---|---|---|---|
| PSNR | 7.2 dB | 8.8 dB | 8.4 dB | **11.7 dB** |
| ZNCC | 0.50 | 0.66 | 0.61 | **0.87** |
| SSIM | 0.29 | 0.37 | 0.31 | **0.43** |

Figure 9: **Polarized NLOS exceeds the quality of conventional NLOS with image processing.** The results of polarized NLOS with partial occluders. The result of the baseline method [32], TV denoised [7], image enhancement by neural network [1], and the result of our method are compared.

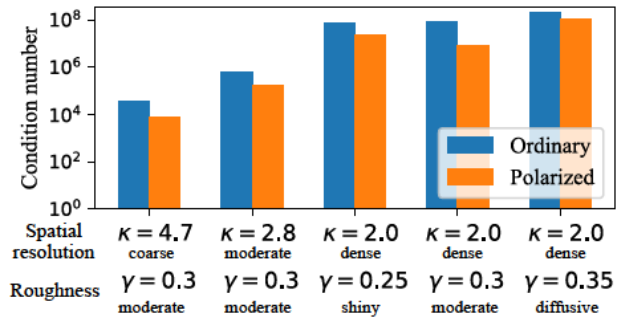

Figure 10: **Condition number of active NLOS.** The condition number of active NLOS setting is compared. The spatial resolution $\kappa$ and the wall's roughness $\gamma$ are changed. Using the polarization cues, the condition number is improved in the active setting for multiple scene configurations.

# Supplementary Material: Enhancing NLOS Imaging Using Polarization Cues

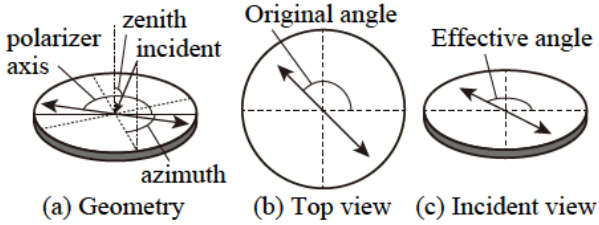

Figure 11: Effective angle of polarizer.

## 8. Effective Angle of Polarizer

Let $a$ and $z$ are azimuth and zenith angles of incident light, respectively, and $\theta$ is the polarizer axis (azimuthial angle) as shown in Fig. 11(a). From the viewpoint of oblique incident light as shown in Fig. 11(c), the vertical axis is shrunk due to the zenith angle of the incident. The cosine and sine of the polarizer axis projected on the waveplane of incident is represented as

$$\begin{cases} \cos\theta' &= \cos(\theta - a + \frac{\pi}{2}) \\ \sin\theta' &= \cos(z)\sin(\theta - a + \frac{\pi}{2}). \end{cases} \quad (16)$$

Therefore, the effective angle $\theta'$ from incident light viewpoint is represented as

$$\tan\theta' = -\frac{\cos(z)}{\tan(\theta - a)}. \quad (17)$$

## 9. Polarized NLOS for Polarized Scene

If the scene is polarized, the observation can be modeled as shown in Eq. (11) in the main text. As the emitted light is completely linear polarized, the reflection off the micro facet is also linearly polarized at any angle. Hence, if the polarizer in front of the camera is placed to block the polarization, the leakage pattern can be clearly observed at any camera position. The modification to the light transport is the same as the general case as shown in Eq. (5) in the main text.

**Experiment** We provide the result for fully polarized scene case (LCD monitor) for reference. For comparison, the scene is recovered with and without polarizer in front of the camera. The camera is put at a non-Brewster angle. Figure 12 shows the results. While the recovered images without polarizer is blurry, recovered images of our method recovers clearly recognizable textures. The numerical evaluation is summarized in Table 2, and we confirm that po-larization cue improves the quality of NLOS imaging for polarized scenes.

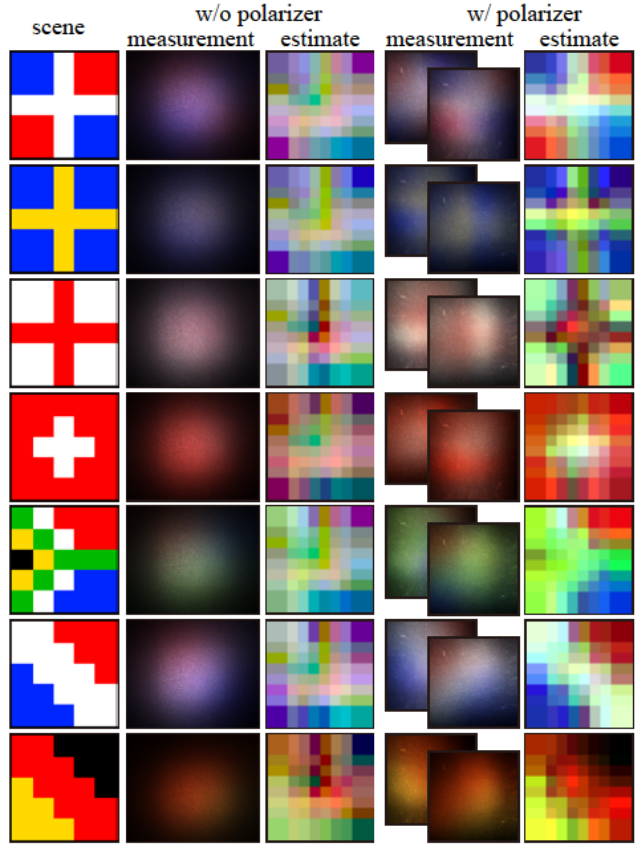

Figure 12: **Polarized NLOS for polarized scene.** The scene is polarized, *i.e.*, LCD monitor. The scene is recovered with and without the polarizer in front of the camera. Using polarization, the recovered images are improved.

|      | PSNR | ZNCC | SSIM |
|------|------|------|------|
| w/o  | 7.3  | 0.40 | 0.30 |
| ours | 10.4 | 0.74 | 0.36 |

Table 2: Numerical evaluations using multiple image measures. Mean values of PSNR, ZNCC, and SSIM are evaluated. For all scene and measure, our method has better value.

## 10. Additional experimental results

**Setup detail** The RGB camera (FLIR BFS-U3-23S3M-C) is looking at the LOS wall only and the scene is placed beyond the occluder wall. The distance between the scene and the wall is approximately 40 cm. For the wall, a black diffusive plate (Thorlabs TPS5) is used and a slice of its BRDF is measured beforehand to construct the light transport matrix. Besides, we assume that the geometry of the wall, the screen, and the camera is known.

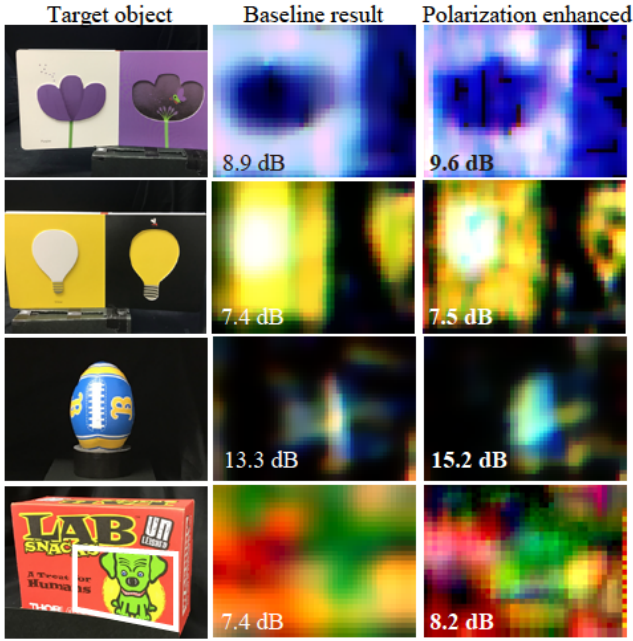

Figure 13: **Results of other reflective scenes.** Our method improves the NLOS imaging. A non-planar scene such as a football is also better recovered than the baseline.

**Polarized NLOS for reflective objects** Figure 13 is other results of reflective objects. Our method can improve not only planar objects such as book pages but also a non-planar objects such as a football. Our diverse results shows the applicability to the general NLOS scenes.

**Images used for numerical evaluations** Figure 14 is the rest of alphabet and flag scenes. The experiment setting is the same as the experiment of the main text. Stand-alone without and with polarizer, and existing method (putting partial obstacle) without and with polarizer are compared. For every scene and setting, the image is improved by using the polarizer.

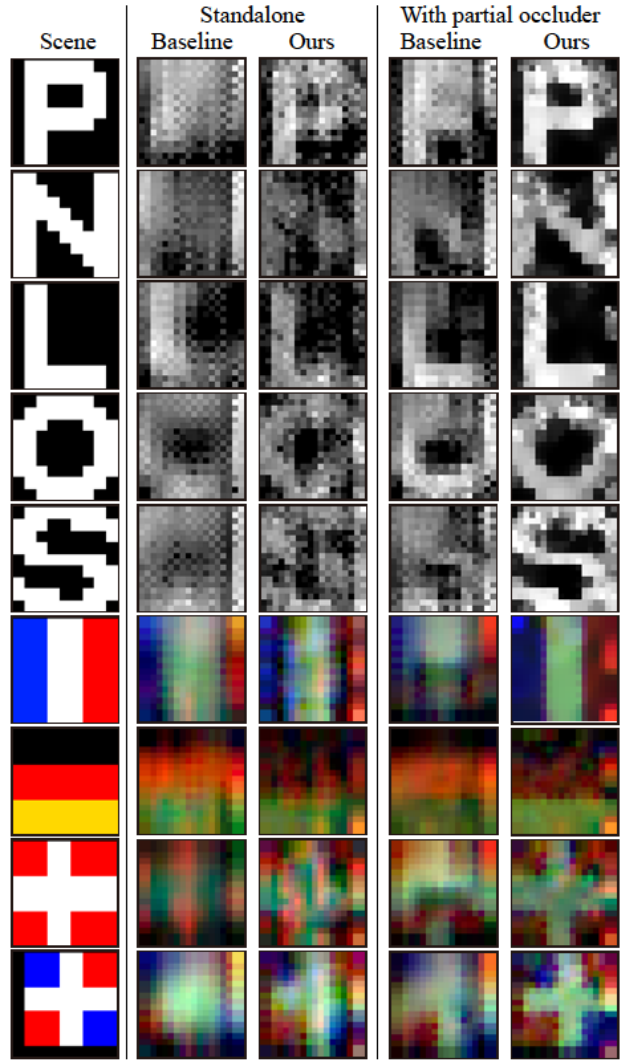

Figure 14: **Polarized NLOS with and without partial occluder.** The scene is projected by a projector and the camera is put at the Brewster angle geometry. The scene is recovered with and without the polarizer. Using polarization cues, the recovered images are improved from the baseline.